 \documentclass[12pt,preprint]{aastex}

\newcommand{\md}{\mbox{d}}
\newcommand{\beq}{\begin{equation}}
\newcommand{\eeq}{\end{equation}}
\newcommand{\beqn}{\begin{eqnarray}}
\newcommand{\eeqn}{\end{eqnarray}}
\newcommand{\lppr}{\stackrel{<}{\scriptstyle \sim}}
\newcommand{\gppr}{\stackrel{>}{\scriptstyle \sim}}

\slugcomment{ApJL accepted}

\shorttitle{On the geometrical origin of periodicity}
\shortauthors{}

\begin{document}


\title{On the geometrical origin of periodicity in blazar-type sources}


\author{Frank M. Rieger
        }
\affil{Department of Mathematical Physics, University College Dublin,
       Belfield, Dublin 4, Ireland}
\email{frank.rieger@ucd.ie}

\begin{abstract}
Periodicities in blazar light curves may be related to helical 
trajectories in extragalactic radio jets by differential Doppler 
boosting effects. We consider ballistic and non-ballistic (i.e., 
radial) trajectories and discuss three possible periodic driving 
mechanisms for the origin of helical jet paths, namely, orbital motion 
in a binary black hole system (BBHS), jet precession, and intrinsic 
jet rotation. It is shown that precessional-driven ballistic motion is 
unlikely to result in observable periods of less than several tens of 
years. We demonstrate that for non-ballistic helical motion the observed 
period is generally strongly shortened relative to the real physical 
driving period because of light-travel time effects. Internal jet rotation 
may thus account for observed periods $P_{\rm obs} \lppr 10$ days. 
Periodicity due to orbital-driven (non-ballistic) helical motion, on the 
other hand, is usually constrained to periods of $P_{\rm obs} \gppr 10$ 
days, while Newtonian-driven precession is unlikely to be responsible for 
periodicity on a timescale $P_{\rm obs} \lppr 100$ days but may well be 
associated with periods of $P_{\rm obs} \gppr 1$ yr.
\end{abstract}


\section{Introduction}
Recent investigations of the light curves in blazar sources have revealed mounting 
evidence of periodicity on different timescales, albeit often with a spread in 
significance: while there seems to be a tendency for a periodicity on a timescale 
of several tens of days in the optical, X-ray or TeV data from TeV sources such
as Mrk~421, Mrk~501, 3C66A and PKS 2155-304 \citep[][]{hay98,lai99,kra01,oso01}, 
the long-term optical light curves from classical sources such as BL Lacertae, 
ON 231, 3C273, OJ~287, PKS 0735+178, 3C345 and AO 0235+16 usually suggest (observed) 
timescales of several years \citep[e.g.,][]{sil88,liu95,fan97,fan98,rai01}. 
Although the significances of the observed periodicities may sometimes be
questioned as they are usually based on limited and uneven sampled data sets, 
their physical possibility is supported by the concurrence of an additional line 
of evidence: high-resolution kinematic studies of parsec-scale radio jets, particularly 
in many of the above noted classical objects, have provided observational evidence 
for the helical motion of components \citep[e.g.,][]{zen88,ste95,vic96,tat98,gom99,
kel04}. A periodically changing viewing angle due to regular helical motion of 
components for example, could via differential Doppler boosting naturally lead to 
periodicities in the observed lightcurves. In this Letter we distinguish 
possible periodic driving mechanisms for helical jet paths by means of their 
associated timescales. Information from observed light curves may thus be used 
in support of ballistic or non-ballistic scenarios and to draw warranted 
conclusions on the nature of the central engine.

\section{Geometrical origin of periodicity and possible driving mechanisms}
For a time-dependent viewing angle, the observed spectral flux modulation by 
Doppler boosting is given by $S_{\nu}(t)=\delta(t)^n\,S_{\nu}'$, with $n = 3 +
\alpha$ for a resolved blob of plasma with spectral index $\alpha$ and $n = 2
+\alpha$ for a continuous flow, $S'$ the spectral flux density in the comoving 
frame, and $\delta$ the Doppler factor defined by
\beq\label{dopplerfactor}
    \delta(t)=\frac{1}{\gamma_{b}(t)\,[1-\beta_b(t)\,\cos\,\theta(t)]}\,,
\eeq where $\theta(t)$ is the actual angle between the velocity $\vec{\beta}_b(t)
=\dot{\vec{x}}_b(t)/c\,$ of the emission region and the direction of the observer, 
and $\gamma_b(t)=1/\sqrt{1-\beta_b(t)^{\,2}}$ is the corresponding bulk Lorentz factor.
A periodically changing viewing angle may thus lead to a periodicity in the observed 
light curve even for an intrinsically constant flux. 

In the case of AGNs, there is well-established evidence for at least three important 
driving mechanisms, possibly resulting in a periodically changing viewing angle: 
(i) non-ballistic helical motion driven by the {\it orbital motion} in a BBHS, 
(ii) ballistic or non-ballistic helical jet paths driven by {\it jet precession} 
and (iii) non-ballistic helical motion due to an {\it internally rotating jet} 
flow. Scenarios following (i) and (ii) usually rely on the plausibility of BBHSs 
in the centres of AGNs. Today, the presence of such systems is indeed 
strongly suggested by hierarchical galaxy evolution schemes \citep[e.g.,][]{beg80} 
and a multitude of observational evidence \citep[e.g.,][for a review]{kom03}.
In the case of simple {\it orbital driving}, a periodically changing viewing angle 
requires non-ballistic (i.e., non-radial) motion of jet elements, so that the orbital 
motion is transmitted to the jet in such a way that the azimuthal velocity of an 
element is characterized by the Keplerian orbital angular frequency 
\beq
  \Omega_k = \sqrt{G\,[M+m]/d^3}\,,
\eeq with $M$ and $m$ the primary and companion mass, respectively, and $d$ the 
separation of the binary. For {\it precessional driving}, on the other hand, both 
ballistic and non-ballistic motion are accompanied by a periodically changing 
viewing angle. There is still some debate on whether the observed helical shapes in 
radio jets are due to ballistic or non-ballistic motion, albeit recent studies 
suggest that non-ballistic motion may be common in many if not most sources 
\citep[][and references in Sect.~1]{kel04}. In general, jet precession may arise
due to relativistic (geodetic and gravitomagnetic) effects, characterized by 
the total angular frequency 
\beq
    \Omega_p^*=(3\,m+\mu)\,G\, \Omega_k/(2\,d\,c^2)\,,
\eeq with $\mu$ the reduced mass, or due to classical effects ({\it Newtonian 
precession}), i.e., tidally induced perturbations in the disk caused by the binary 
companion and resulting in a rigid body precession of the inner parts of the disk
\citep[][]{kat97,lar98,rom00,rom03}.
Non-ballistic helical motion of knots with periodically changing viewing angles may, 
however, also be a natural consequence of an overall straight yet {\it internally 
rotating jet} (scenario iii), characterized by the angular frequency $\Omega(r)$, 
provided that knots are dragged with the rotating flow. Again, such a scenario appears 
well-supported given the evidence for a close link between the jet and the disk \citep[e.g.,]
[]{raw91,fal95}, suggesting that a significant amount of rotational energy of the disk 
is channeled into the jet.

\section{Travel-time effects for non-ballistic helical motion}
In order to correctly extract the relevant information such as the timescale of 
periodicity from measured light curves, light-travel time effects for non-ballistic 
helical motion need to be taken into account. 
For simplicity, consider an emitting component which moves relativistically along an 
idealized helical path at small inclination toward the observer.
Denote by $\vec i = (\sin i,\,0,\, \cos i)$ the normalized vector pointing toward 
the observer and let $\vec x_b(t)=(R\,\cos \Omega\,t,\,R\,\sin \Omega\,t\,,\,v_z\,t)$
be the position vector of the emission region, with $v_z$ the outflow velocity along 
the $z$-axis. For simple, {\it orbital-driven} helical motion (scenario i), $R \equiv
M\,d/(m+M)$, $\Omega \equiv \Omega_k$, and $i$ is the angle between the direction of 
the total angular momentum and the line-of-sight (LOS). 
{\it Precessional driving} (scenario ii), on the other hand, may be modelled by setting 
$R \equiv R(t) = v_z\,t\,\tan\alpha$, with $\alpha \leq i$ the half opening angle of the 
precession cone, $\Omega \equiv \Omega_p$ and with $i$ the angle between the cone axis 
and LOS. Finally, for a well-collimated, {\it internally rotating jet} flow (scenario 
iii), $R$ may be identified with the radial scale $r_0$ at which the knot is injected, 
$\Omega \equiv \Omega(r_0)$ and $i$ is the angle between the jet axis and LOS.\\
In general, we measure a peak in the observed flux each time the velocity 
vector of the component points closest toward the observer, say at point~A 
and point~B. In the frame fixed in the center of the galaxy, the component moves from 
point~A to point~B within the time span $P=2\,\pi/\Omega$. Technically, the observed 
period is determined by measuring the arrival times of light pulses emitted at A and B. 
This period however, will usually be much shorter than the (real) physical period $P$, 
as the travel distance for a pulse emitted at B is much shorter than the one for a pulse 
emitted at A, due to the relativistic motion of the component in the direction of the 
observer. The observed difference in arrival times depends on the projection of the 
velocity vector of the knot $\vec{\beta}_b(t) = \dot{\vec x}_b(t)/c$ on the direction 
$\vec{i}$ to the observer. Hence, for an infinitesimal time interval $\md t$ we have
\begin{equation}
     \md t_{\rm obs} = (1+z)\,[1-\vec{\beta}_b(t) \cdot \vec{i}\,] \,\md t
                     = (1+z)\,\left[1-\beta_b(t)\,\cos \theta(t)\,\right]\,\md t\,,
\end{equation} where the factor $(1+z)$ accounts for a possible redshift 
dependence and $\cos \theta(t)=\vec i \cdot \dot{\vec x}_b/(|\vec i|\,|
\dot{\vec x}_b|)$. For the relation between the observed period $P_{\rm obs}$ 
and $P$ one thus obtains
\beq
     P_{\rm obs}  =  (1+z)\,\int_0^P\,\,\left[1-\beta_b(t)\,
                           \cos\theta(t)\,\right]\; \md t\,\,.
\eeq 
Performing the integration in the case of {\it orbital-driven} helical motion 
one finds \citep[][]{rie00}
\beqn\label{orbital}
  P_{\rm obs} & = & (1+z) \,\left[1-\frac{v_z}{c}\,\cos i\,\right]\,P_k\,,
\eeqn assuming that the observed emission is dominated by radiation from
a single jet. The real period may be twice as long if both black holes (BHs) have 
similar jets. A similar dependency is found for an {\it internally rotating} 
flow, i.e. $P_{\rm obs} = (1+z)\,[1 - (v_z/c)\,\cos i]\,P(r_0)$ \citep[cf. 
also][]{cam92}. For {\it precessional driving}, on the other hand, one has
\beq 
  P_{\rm obs} = (1+z) \,\left[1-\frac{v_z}{c}\,\cos i
                      -\frac{v_z}{c}\tan\alpha\,\sin i\right]\,P_p\,,
\eeq where the $\alpha$-dependence is due to motion along the surface of the 
cone.

Now, for blazar-type sources we may obtain a representative scale for the 
(real) physical driving period by using classical parameters inferred from 
high energy emission models, i.e., inclination angles $i \simeq 1/\gamma_b$ 
and bulk Lorentz factors $\gamma_b \simeq (5-15)$ \citep[e.g.,][]{chi00}, 
which yields
\beq\label{approxi}
    P  \simeq \, \frac{\gamma_b^2}{(1+z)}\,\,\,P_{\rm obs} \,= \frac{13.7}{(1+z)}
                 \,\,\left(\frac{\gamma_b}{10}\right)^2\,
                  \left(\frac{P_{\rm obs}}{50\, {\rm days}}\right)\;{\rm yr}\,
\eeq assuming non-relativistic rotational velocities, i.e. $v_z/c \simeq 
(1-1/\gamma_b^2)^{1/2}$. The above estimate is also valid for precessional 
driving, provided that the half-opening angle $\alpha$ is small enough. 
Note, however, that the situation is different for a ballistic helical-type jet, 
where individual fluid elements move radially away from the origin, with the 
direction of ejection changing periodically with time due to precession. 
The origin of periodicity in this case could be related to a periodically new 
and dominant ejection from the centre (i.e. not due to the motion of a single 
element or component), with the older ones fading away with distance (e.g. due 
to an inhomogeneous-type flow or deceleration). There is thus no shortening of 
the period in such cases, apart from the usual redshift correction, since all 
relevant signals essentially do have to travel the same distance toward the 
observer. In general a driving period $P$ translates into a characteristic 
projected length scale (wavelength) for the associated helical trajectory of 
$\lambda = P\,v_z\,\sin i$, i.e.
\beq
  \lambda_{\rm nb} \;\simeq\; P_{\rm obs}\,c\,\gamma_b\,/\,(1+z)
  \quad {\rm and} \quad \lambda_{\rm b} \;\simeq \;\lambda_{\rm nb}\,/\,\gamma_b^2\,
\eeq with $\lambda_{\rm b}$ for ballistic and $\lambda_{\rm nb}$ for non-ballistic 
motion, e.g., $\lambda$ being much smaller for a ballistic than for a non-ballistic
origin of periodicity.

\section{Constraints and associated timescales for specific driving mechanisms}
In general, the above noted driving mechanisms may be distinguished according
to their associated timescales:

Periodicity caused by {\it orbital-driven} non-ballistic helical motion for 
example, probably represents the scenario with the smallest degrees of 
freedom. Existing supermassive binary models aimed at explaining periodicity 
in blazars by orbital or precessional motion usually require very close BBHSs 
with typical separations of the order $d \simeq 5 \cdot 10^{16}\,(P_k/10\,
{\rm yr})^{2/3}\,([M+m]/[5\cdot 10^8\,M_{\odot}])^{1/3}$ cm and orbital periods 
of several years, suggesting observable periods $\gppr 10$ days. In general, 
such close binaries are quite vulnerable to quickly lose their orbital angular 
momentum via gravitational radiation. Stability of the observed period or 
arguments from cosmological evolution may thus be used for individual sources 
to impose a lower limit on the required lifetime and hence the separation and 
orbital period of the putative BBHS \citep[e.g.,][]{rie00}. Note that for 
orbital-driven helical motion the periodicity at a given frequency is expected 
to be disturbed or even washed out, when the width of the jet (at that frequency!) 
becomes comparable to the size of the orbit, which naturally limits the number 
of observable periods.

Periodicity caused by {\it precessional-driven} helical motion is normally
associated with driving periods much higher than the orbital period. While
a geodetic and gravitomagnetic precessional origin of periodicity may be
neglected by virtue of the associated long driving periods (usually in excess 
of $P_p \sim 10^4$ yr), Newtonian-driven jet precession caused by rigid body 
precession of the inner parts of the disk may represent a plausible option. 
It could be shown that for such a scenario the ratio between the orbital 
($P_k$) and precessional ($P_p$) period obeys \citep[cf.][]{lar98,rom03}  
\beq\label{ratio_periods}
       \frac{P_k}{P_p} \simeq \frac{3}{7}\,\frac{m}{M}\,
                        \left(\frac{r_d}{d}\right)^{3/2}\,
                        \left(\frac{M}{m+M}\right)^{1/2}\,\cos\alpha\,\,,
\eeq where $m$ is the mass of the BH exerting the gravitational torque, $d$ is
the orbital separation of the binary, $r_d$ is the outer radius of the precessing 
inner disk, with $r_d/d < 1$ and $\alpha$ the precession cone half-angle. 
Newtonian (rigid body) precession requires that induced bending disturbances
take on the character of propagating waves, a behaviour expected to occur for 
$\alpha < 1/{\cal M}$, where ${\cal M}$ denotes the mean Mach number of the 
disk and $\alpha$ the usual dimensionless kinematic viscosity. 
Differential precession may then be smoothed out if the crossing time for the 
bending waves, propagating at nearly one third of the sound speed $c_s$, is much 
shorter than the precessional period \citep[][]{papt95,lar98}. Hence the condition 
for rigid body rotation becomes $c_s \gg 3\,\Omega_{\rm prec}\,r_d$ \citep[][]{lar98}. 
Combining these relations, one obtains $P_k/P_p \,\ll\,1/(7\,{\cal M})^{1/2}$, 
implying a characteristic lower bound for the precessional period in a close 
BBHS of
\beq
      P_p \,\gg\, 40\,\,\left(\frac{\cal M}{10}\right)^{1/2}
                \left(\frac{P_k}{5\,{\rm yr}}\right)\;\;{\rm yr}\,
\eeq for a typical Mach number ${\cal M} \sim 10$. This suggests (a) that 
ballistic motion is generally unable to account plausibly for periodicities with 
observed periods less than several tens of years and (b) that unless very high
flow Lorentz factors are invoked, non-ballistic Newtonian-driven precession in a 
close BBHS is unlikely to result in observable periods of less than $100$ days.
However, periods of the order of several years, as observed for example in the 
BL Lac object AO~0235+16, may be generally due to (Newtonian) precessional-driven 
as well as orbital-driven non-ballistic helical jet paths \citep[cf. also,][]{rom03,
ost04}.

In the above mentioned scenarios, the physical evolution of the non-ballistic 
helical path is usually treated somewhat heuristically. What in any case is 
required in such models is a coherent jet channel with some kind of rigidity 
or stiffness, e.g. as a consequence of the flow being wrapped by magnetic fields 
\citep[e.g.,][]{con93,fenz98}. Note that periodic variability may also arise if jets 
are characterized by a global helical magnetic field. A more detailed analysis of 
such a possibility \citep[cf. also,][]{rol94} will be given elsewhere.

In the case of periodicity driven by an {\it internally rotating flow}, a 
physically more detailed scenario, based on MHD jet models, has been proposed by 
\citet[][]{cam92} and applied to several sources \citep[e.g.,][]{wag95}.
Assuming that the magnetic flux giving rise to the jet is concentrated toward 
the innermost part of the disk, the origin of periodicity has been related to 
an off-axisymmetric knot, with a size much smaller than the jet radius $r_j$, 
which is injected at radius $r_0$ beyond the light cylinder $r_{\rm L} \simeq 1.5 
\cdot 10^{14}\,M_8$ cm (with $M_8$ the BH mass in units of $10^8\,M_{\odot}$) 
and dragged with the underlying rotating flow. As the plasma rotation beyond 
$r_{\rm L}$ is considered to be governed by specific angular momentum conservation, 
a given period $P_{\rm obs}$ may be related to a specific injection point $r_0$ 
in the almost cylindrically-collimated jet region by (cf. Eq.~\ref{approxi}, but 
allowing for sub-relativistic rotational velocities of order $\sim 0.1$ c)
\beqn
   r_0 & \simeq & \sqrt{\frac{c}{2\,\pi}}\; r_{\rm L}^{1/2}\;P^{1/2} \\
       & \simeq &  \frac{120}{\sqrt{1+z}}\,\left(\frac{\gamma_b}{10}\right)\,
       \left[1 +\frac{1}{2}\,\left(\frac{\gamma_b}{10}\right)^2
       \left(\frac{v_{\theta}(r_0)}{0.1\,c}\right)^2\right]^{-1/2}
       \frac{1}{\sqrt{M_8}}\;
       \left(\frac{P_{\rm obs}}{50\, {\rm days}}\right)^{1/2} r_{\rm L}\,.
       \nonumber
\eeqn 
Numerical MHD simulations have shown, that the asymptotic, cylindrical jet radius 
in these models is usually confined to $r_{\rm L} < r_0 < 10\,r_{\rm L}$ \citep[e.g.,]
[]{cam96,fen97}. This suggests, that unless very low flow Lorentz factors and/or 
unusual high BH masses are invoked, consistency implies that the observed timescale of 
periodicity for BL Lac-type objects due to internal jet rotation is generally limited 
to $P_{\rm obs} \sim 1$ day or less, while for the more massive quasars one may have 
$P_{\rm obs} \lppr 10$ days. In general, almost perfect collimation with an intrinsic 
opening angle of less than $0.1^{\circ}$ is required in such scenarios, as otherwise 
the periodicity will be quickly washed out due to the requirement of angular momentum 
conservation. Again, this could be used to impose an upper limit on the possible number 
$N$ of observable periods as the jet usually needs to open before entering the pc-scale.
Note, however, that the real situation may be more complex if other rotation profiles 
\citep[e.g.,][]{vla98} are indeed realized.

\section{Consequences}
In the case of blazar-type sources where most of the observed flux is usually 
dominated by non-thermal emission from their relativistic jets, periodic variability 
may occur as a result of differential Doppler boosting associated with helical jet paths. 
Whereas this conclusion appears physically robust, the detectability of the associated 
period in a given source will depend on several conditions, e.g. whether the periodicity 
at a given frequency is dominated by a strong disk contribution (in the optical-UV), 
affected by absorption (e.g., $\alpha =-2.5$ in the radio regime), or washed out due 
to internal jet stratification.

Periodicity by differential Doppler boosting may arise due to precessional-driven ballistic 
motion of components or due to non-ballistic helical motion driven by (i) the orbital motion 
or (ii) Newtonian precession in a close BBHS, or (iii) internal jet rotation. We have 
demonstrated that for non-ballistic helical motion the observed period $P_{\rm obs}$ may 
be much smaller than the underlying physical driving period $P_d$, i.e. $P_{\rm obs} \sim 
(1+z)\,P_d/\gamma_b^2$. This result could be utilized for assessing the plausibility of 
ballistic or non-ballistic motion in a given radio jet. For example, a closer inspection 
reveals that precessional-driven ballistic motion is unlikely to be associated
with observed periods of less than several tens of years. Conversely, observed 
periods less than several years provide strong evidence for a non-ballistic origin. 
Such non-ballistic scenarios may further be distinguished according to their inherent
constraints: internal jet rotation in a lighthouse-type manner \citep[][]{cam92}, for 
example, is expected to result in $P_{\rm obs} \lppr 10$ days (for massive quasars), with
periods likely to be of order $P_{\rm obs}\sim 1$ day for BL Lac objects. Periodicity due 
to orbital-driven helical motion \citep[][]{rie00} is usually constrained to periods 
of $P_{\rm obs} \gppr 10$ days and for massive binaries expected to be well above several 
tens of days. Newtonian-driven precession on the other hand, seems unable to account 
plausibly for periods $P_{\rm obs} \lppr 100$ d, but may well be associated with $P_{\rm 
obs} \gppr 1$ yr. This suggests, for example, that periodic variability with timescales 
of several tens of days, as apparently observed in some TeV sources, is most likely 
caused by orbital-driven helical motion. Moreover, if the jet evolution in a BBHS is 
sufficiently inhomogeneous, the high-energy emission will only evince effects of 
orbital modulation while precessional modulation may be present in lower energy bands.

\acknowledgments
Discussions with G.~Romero, J.-H.~Fan, and K.~Mannheim, and support by a 
Marie-Curie Individual Fellowship (MCIF-2002-00842) are gratefully 
acknowledged.

\end{document}